\documentclass[aps,pra,twocolumn,showpacs,floatfix,superscriptaddress]{revtex4}
\usepackage{color}      % use if color is used in text
\usepackage{amsmath,amssymb,amsfonts,bbm,graphicx,hyperref,times}
\newcommand{\phiT}{\tilde{\phi}}
\newcommand{\U}{\Upsilon}

\newcommand{\be}{\begin{equation}}
\newcommand{\ee}{\end{equation}}
\newcommand{\bea}{\begin{eqnarray}}
\newcommand{\eea}{\end{eqnarray}}

\def\Tr{\hbox{Tr}} 
\def\sigmaCM{\boldsymbol{\sigma}}
\begin{document}
%%%%%%%%%%%%%%%%%%%%%%%%%%%%%%%%%
\title{General-dyne unravelling of a thermal master equation}
%\date{\today}
%%%%%%%%%%%%%%%%%%%%%%%%%%%%%%%%%%%%%%%%%
\author{Marco G. Genoni}
\affiliation{Department of Physics \& Astronomy, University College London, 
Gower Street, London WC1E 6BT, United Kingdom}
\email{marco.genoni@ucl.ac.uk}
\author{Stefano Mancini}
\affiliation{School of Science and Technology, University of Camerino, I-62032 Camerino, Italy \\
and INFN, Sezione di Perugia, I-06123 Perugia, Italy}
%\email{stefano.mancini@unicam.it}
\author{Alessio Serafini}
\affiliation{Department of Physics \& Astronomy, University College London, 
Gower Street, London WC1E 6BT, United Kingdom}
\affiliation{Scuola Normale Superiore, I-56126 Pisa, Italy}
%\email{serale@theory.phys.ucl.ac.uk}

%%%%%%%%%%%%%%%%%%%%%%%%%%%%%%%%%
\begin{abstract}
We analyse the unravelling of the quantum optical master equation at finite temperature due to direct, continuous, general-dyne detection of the environment. We first express the general-dyne Positive Operator Valued Measure (POVM) in terms of the eigenstates of a non-hermitian operator associated to the general-dyne measurement. Then, we derive the stochastic master equation obtained by considering the interaction between the system and a reservoir at thermal equilibrium, which is measured according to the POVM previously determined. Finally, we present a feasible measurement scheme which reproduces general-dyne detection for any value of the parameter characterising the stochastic master equation.
\end{abstract}
\maketitle
\section{Introduction}

An expedient characterisation and classification of quantum measurements is central to quantum control tasks \cite{WisemanMilburn}, especially when the optimisation of certain figures of merit is involved \cite{opt2,opt3,opt4,boundTH,opt5}. In the context of the coherent control of quantum continuous variables, consistently pursued, over the last thirty years, since early works by Belavkin \cite{BelavFiltering1,BelavFiltering2,Belavkin3}, the class of general-dyne measurements stands out as it is associated to all diffusive unravellings of the dynamics, {\em i.e.} to all the unravellings that can be treated as multivariate quantum Wiener processes \cite{Barchielli, WisemanDiosi,WisemanDoherty}. 
Such conditional dynamics share the property of preserving the Gaussian nature of the system's state, and thus allow for an extensive analytical treatment. 
Let us remind the reader that general-dyne measurements include the well known homodyne and heterodyne detection schemes as special cases.
In this paper, we consider a system of one degree of freedom coupled with a thermal reservoir at non-zero temperature, and derive the unravelling of the 
master equation enacted by general-dyne detection on the environmental degree of freedom coupled to the system. Further, we identify a feasible 
measurement scheme to perform general-dyne detection, and relate it explicitly to the parameter that specifies the general-dyne unravelling of the master 
equation.

Note that our treatment holds at finite, non-zero temperature, which is particularly relevant to mechanical systems where the optical master equation still applies. The cooling and coherent control of such systems is now very much in the limelight of research in experimental opto-mechanics and quantum optics \cite{OptoMech1,OptoMech2,cooling,FBAlessio}.

The paper is structured as follows: In Sec. \ref{s:thermal} we present the stochastic unreavelling of non-zero temperature master equation for a single bosonic mode: specifically, in Sec. \ref{s:eigenstate} we derive the 
{Positive Operator Valued Measure (POVM)} of the measurement associated with the {general-dyne operator $\Theta$}, while in Sec. \ref{s:singleunravel} we derive the corresponding stochastic master equation (SME). 
Finally, in Sec. \ref{s:scheme}, we describe a measurement scheme able to measure the general-dyne observable $\Theta$. We end the paper in Sec. \ref{s:conclusions} with some concluding remarks and outlook.

\section{Thermal master equation and general-dyne stochastic unravellings} \label{s:thermal}
We consider here a quantum harmonic oscillator described by bosonic operators $[c,c^\dagger]=\mathbbm{1}$,  interacting 
with a non-zero temperature bath with an average thermal photon number $N$.
The corresponding time evolution is described by the Lindblad Master Equation {(see, {\em e.g.}, \cite{WisemanMilburn})}
\begin{align}
d\varrho &= \mathcal{L}_{\sf th} \varrho \: dt \nonumber \\
&= (N+1) \mathcal{D}[c] \varrho \: dt + N \mathcal{D}[c^\dag] \varrho \: dt \label{eq:ME}
\end{align}
where $\mathcal{D}[O]\varrho = O\varrho O^\dag - (O^\dag O \varrho + \varrho O^\dag O)/2$.
We assume to monitor continuously the environment on time scales which are much
shorter than the typical system's response time, by means of
weak measurements. The dynamics will be then described by a SME, depending on the type of measurement performed on the bath. In this manuscript we will consider a general-dyne measurement, which will be introduced in the next section, along with the corresponding POVM.
\subsection{General-dyne POVM} \label{s:eigenstate}
{Given a field mode described  by bosonic operators $[a,a^\dagger]=\mathbbm{1}$, 
the} so-called general-dyne detection corresponds to the monitoring of the following (non-Hermitian) operator
\begin{equation}
\Theta=a+\Upsilon a^{\dag}, \label{eq:Theta}
\end{equation} 
where, for the sake of simplicity, we assume $\Upsilon\in[-1,1]\subset\mathbb{R}$; in the extreme cases $\Upsilon=\pm 1$ it corresponds to the homodyne detection of respectively the {'position' and 'momentum' } quadratures $q$ and $p$, while for $\Upsilon=0$ it corresponds to heterodyne detection. 

In order to derive the eigenstates of $\Theta$ we can use the eigenvalues equation
\begin{equation}
a+\Upsilon a^{\dag} |\theta\rangle=\theta |\theta\rangle, 
\label{eqeval}
\end{equation}
where w.l.g. $\theta\in\mathbb{C}$, {and
we} consider the canonical position and momentum operators defined by
\begin{eqnarray}
q&=&a+a^{\dag},\\
p&=&-i(a-a^{\dag}),
\end{eqnarray}
with commutation relation $[q,p]=2i$.
Letting $|x\rangle$ (resp. $x$) be an ``improper'' eigenvector (resp. eigenvalue) of $q$, we have the following correspondences
\begin{eqnarray}
q &\leftrightarrow& x,\\
p &\leftrightarrow& -2i\frac{d}{dx}.
\end{eqnarray}
In view of that we can turn \eqref{eqeval}
into an ordinary differential equation for $\psi(x)=\langle x|\theta\rangle$, namely
\begin{equation}
\frac{1}{2}\left[(1+\Upsilon)x+2(1-\Upsilon)\frac{d}{dx}\right]\psi(x)=\theta \psi(x).
\label{ev}
\end{equation}
Solving this equation we obtain 
\begin{align}
\psi(x)&=\left[\frac{1}{2\pi}\frac{1+\Upsilon}{1-\Upsilon}\right]^{1/4} \times \nonumber \\
& \times \exp\left[-\frac{1}{2}\left(\sqrt{\frac{1+\Upsilon}{2(1-\Upsilon)}}\;x-\sqrt{\frac{2}{(1-\Upsilon^2)}}\;\theta_1\right)^2 + \right. \nonumber \\ 
&\:\:\:\: \left. +i\frac{\theta_2} {1-\Upsilon}\; x\right],
\end{align}
where the subscript $1$ (resp. $2$) indicates the real (resp. imaginary) part.
Clearly it is
\begin{equation}
|\theta\rangle=\int dx \: \psi(x) |x\rangle.
\label{thx}
\end{equation}
By using this equation and the completeness relation for the position eigenstates $|x\rangle$, we notice that the POVM corresponding to the measurement of the operator $\Theta$, with outcome $\theta$, 
is given by 
\begin{equation}
d\Pi(\theta)=|\theta\rangle\langle\theta| \frac{d\theta_1 d\theta_2}{\pi(1-\Upsilon^2)}. \label{eq:POVM}
\end{equation}

\subsection{General-dyne stochastic master equation} \label{s:singleunravel}
The master equation (\ref{eq:ME}) is obtained by considering a 
harmonic oscillator interacting by a beam splitting interaction with a bath mode in a 
thermal state at non-zero temperature, with $N$ thermal photons 
on average. 
Let us consider at time $t$ the quantum state $R(t)=\varrho(t)\otimes \mu(t)$, where 
$\varrho(t)$ and $\mu(t)$ represent respectively the state of the system and 
of the bath.  
In order to describe the effect due to a continuous measurement of the bath, we will
follow the procedure used in Ref. \cite{WisemanThesis}. We start by transforming 
the bath state into a Wigner probability distribution obtaining the operator (in the
system Hilbert space)
\begin{align} 
%R(t) \rightarrow 
\widetilde{W}(t) &= \int  \frac{d^2 \lambda }{\pi^2} {\rm Tr}_{\sf B} \left[ R\: e^{\left\{\lambda_1 (a^{\dag}-\alpha^*)-\lambda_1^*(a-\alpha) \right\}}\right] \\
&= W_t(\alpha) \varrho(t) .
\end{align}
where
\begin{align}
W_t(\alpha) &= \frac{1}{\pi (N+\frac{1}{2})}\exp\left[-\frac{|\alpha|^2}{(N+\frac{1}{2})}\right] \:,
\label{WN}
\end{align}
denotes the Wigner function of a (single-mode) thermal state with $N$ thermal photons.

Notice that above we have introduced the bosonic operator $a=\sqrt{dt} b(t)$, satisfying the commutation relation $[a,a^\dag]=\mathbbm{1}$, while the operator $b(t)$, which describes the reservoir with infinite bandwidth, satisfies $[b(t),b^\dag(t')] = \delta(t-t')$. 
After an infinitesimal time $dt$ the state describing system and the bath evolves as 
\begin{align}
R(t+dt) &= R(t) + dt [ b^\dag(t) c-c^\dag b(t), R(t)] + O(dt)  \\
&= R(t) + \sqrt{dt} [ a^\dag c-c^\dag a, R(t)] + O(dt) \:,
\end{align}
which in the Wigner function picture reads
\begin{align}
\widetilde{W}(t+dt) &= \widetilde{W}(t) + \sqrt{dt}\left[ (\alpha^* -\frac12 \partial_\alpha) c \widetilde{W}(t) - \right. \nonumber \\
& \: \left. - (\alpha+\frac12 \partial_{\alpha^*})c^\dag \widetilde{W}(t) - (\alpha^*+\frac12\partial_\alpha) \widetilde{W}(t) c + 
\right. \nonumber \\ 
&\:\: + \left. (\alpha-\frac12 \partial_{\alpha^*}) \widetilde{W}(t) c^\dag \right] + O(dt) \label{eq:wignerevolv} \:.
\end{align}
We then consider a continuous measurement of the observable $\Theta$,  described by the POVM 
(\ref{eq:POVM}), yielding the (unnormalised) conditional state
\begin{align}
\widetilde{W}_c(t) &= \frac{\langle \theta | R(t+dt) |\theta \rangle}{\pi (1-\Upsilon^2)} \nonumber \\
&= \int d^2\alpha\: dx \: dx' \: \widetilde{W}(t+dt) \times \nonumber \\
&\:\: \times  \psi^*(x)\psi(x')e^{i\alpha_2(x-x')}\delta\left(2\alpha_1-\frac{x+x'}{2}\right).
\end{align}
By performing the derivatives and the integrals, we obtain
\begin{widetext}
\begin{align}
\widetilde{W}_c(t) &= p(\theta_1,\theta_2 ; t) \left\{ \varrho(t) + \sqrt{dt} \frac{\theta_1}{1+N(1+\Upsilon)} \left[ (N+1) c \varrho(t) - N c^\dag \varrho(t) + (N+1)\varrho c^\dag - N \varrho(t) c \right] \right.+ \nonumber \\
&\left. \:\:\:\: -i \sqrt{dt} \frac{\theta_2}{1+N(1+\Upsilon)} \left[ (N+1) c \varrho(t) - N c^\dag \varrho(t) + (N+1)\varrho c^\dag - N \varrho(t) c \right] 
\right\} \:.
\end{align}
\end{widetext}
The  outcomes probability $p(\theta_1,\theta_2 ; t)$ at time $t$ is a zero-centered Gaussian function with covariance matrix $\sigma={\rm diag}(L_1,L_2)$, 
%\begin{align}
%\sigma = 
%\left( 
%\begin{array}{c c}
%L_1 & 0 \\
%0 & L_2
%\end{array}
%\right) ,
%\end{align}
where $L_{1/2} = (1\pm\Upsilon)(1+N(1\pm \Upsilon))/2$. 
One can easily check that for $\Upsilon= 1$ one has 
\begin{equation}
p(\theta_1,\theta_2; t)=\sqrt{\frac{1}{2\pi(1+2N)}}
\exp\left[ -\frac{\theta_1^2}{2(1+2N)}
\right]\delta(\theta_2),
\end{equation}
which corresponds to the probability distribution for the measurement of the $q$ quadrature of
a thermal state.
Likewise, for $\Upsilon=0$ one obtains 
\begin{equation}
p(\theta_1,\theta_2; t)=\frac{1}{\pi(1+N)}
\exp\left[ -\frac{\theta_1^2}{1+N}
-\frac{\theta_2^2}{1+N}
\right],
\end{equation}
which corresponds to the Husimi-Q function of the thermal state {\cite{S01}}, and thus to the probability distribution for etherodyne detection on the bath mode.

By calculating the trace of the conditional state we obtain the probability of obtaining the result $\theta=\theta_1+i\theta_2$ from the measurement at time $t+dt$,
\begin{widetext}
\begin{align}
p(\theta_1,\theta_2; t+dt) &= \hbox{Tr}_s [ \widetilde{W}_c(t+dt) ] \\
&= p(\theta_1,\theta_2; t)\left\{ 1+ \frac{\sqrt{dt}}{L_1}\frac{1+\Upsilon}{2} \langle c +c^\dag\rangle + \frac{\sqrt{dt}}{L_2}\frac{1-\Upsilon}{2} \langle i(c^\dag-c)\rangle \right\} + O(dt) \:.
\end{align}
\end{widetext}
This result allows us to consider the two variables $\theta_j$ as Gaussian random variables
\begin{align}
\sqrt{dt} \theta_1 &= \frac{1+\Upsilon}{2} \langle c + c^\dag \rangle(t) dt + \sqrt{L_1}\: dw_1 \label{eq:dw1}\\
\sqrt{dt} \theta_1 &= \frac{1-\Upsilon}{2} \langle i(c^\dag-c) \rangle(t) dt + \sqrt{L_2}\: dw_2 \label{eq:dw2}
\end{align}
where we define the uncorrelated Wiener increments s.t. $dw_j^2=dt$ and $dw_1 dw_2=0$.
After having obtained the normalized conditional state by using the formula
\begin{align}
\varrho_c (t+dt) &= \frac{\widetilde{W}_c(t+dt)}{p(\theta_1,\theta_2;t+dt)} \:,
\end{align}
%{\red [What are $q_A$ and $q_B$?]} 
and by using the relations (\ref{eq:dw1}) and (\ref{eq:dw2}), one obtains the SME
\begin{widetext}
\begin{align}
d\varrho_c(t) &= \varrho_c (t+dt)  - \varrho_c (t) \nonumber \\
&= \frac{1+\Upsilon}{2} \frac{dw_1}{\sqrt{L_1}}
\mathcal{H}[(N+1) c - N c^\dagger] \varrho_c(t) +  \frac{1-\Upsilon}{2} \frac{dw_2}{\sqrt{L_2}} \mathcal{H}[(N+1)(-ic) - N (ic^\dagger)]
\varrho_c(t) + O(dt) \:. \label{eq:genericSME}
\end{align}
\end{widetext}
The equation reported here is the most general unravelling of the thermal master equation, corresponding to general-dyne detection of the mixed thermal bath, if no (partial or total) purification of the bath is accessible.
By setting $\Upsilon=1$ we obtain the SME corresponding to continuous homodyne measurement
of the $q$ quadrature of a thermal bath \cite{WisemanMilburn}, 
\begin{align}
d\varrho_c(t) &=  \frac{dw_1}{\sqrt{(1+2N)}}\mathcal{H}[(N+1) c - N c^\dagger] \varrho_c(t)  + O(dt) \:.
\end{align}
Analogously, by setting $\Upsilon=0$ we obtain the SME corresponding to continuous etherodyne detection,
\begin{align}
d\varrho_c(t) &= 
\mathcal{H}[(N+1) c - N c^\dagger] \varrho_c(t) \frac{dw_1}{\sqrt{2(1+N)}} + \nonumber \\
&\mathcal{H}[(N+1)(-ic) - N (ic^\dagger)]\varrho_c(t)\frac{dw_2}{\sqrt{2 (1+N)}} + O(dt) .
\end{align}
%\textcolor{red}{{\sf Shall we write that the SME for etherodyne in Wiseman's thesis is wrong? Nah...let's just give the right form above}}\\
This general SME can be also used to investigate the usefulness of continuous-measurement (and feedback) for practical purposes, as the generation of continuous-variable entanglement and squeezing.
In \cite{boundTH} it was proven that high value of squeezing could be obtained by means of continuous measurement and feedback for a system whose evolution is described by the thermal master equation (\ref{eq:ME}); in particular the bound on the variance for a certain quadrature reads
\begin{align}
\Delta x^2 \geq \frac 1 {2N+1},
\end{align}
which decreases, and thus yields a larger degree of squeezing, by increasing the temperature of the bath.
However one can prove that, by varying $\Upsilon$ in the whole range of values $[-1,1]$ the bound on the achievable squeezing cannot be saturated and the achieved steady state is always a thermal state with $N$ average photons. This clearly shows that, when the bath is in a mixed state, one cannot achieve the ultimate bound on squeezing (and entanglement, in the multipartite case), by direct general-dyne detection of the 
environmental degrees of freedom. More general Gaussian measurements, obtained by adding entangled ancillary modes, are in order to that aim. 
%one needs to resolve it in its pure states decomposition, {\em i.e.}, as shown in Ref. \cite{Ravotto}, one has to consider its purification and perform continuous monitoring on the corresponding overall pure (bipartite) state.
% {\red [I would remove the last sentence, starting from i.e., and Ref. \cite{Ravotto}]}
   %
\begin{figure}[h!]
\resizebox{.90\columnwidth}{!}{\includegraphics{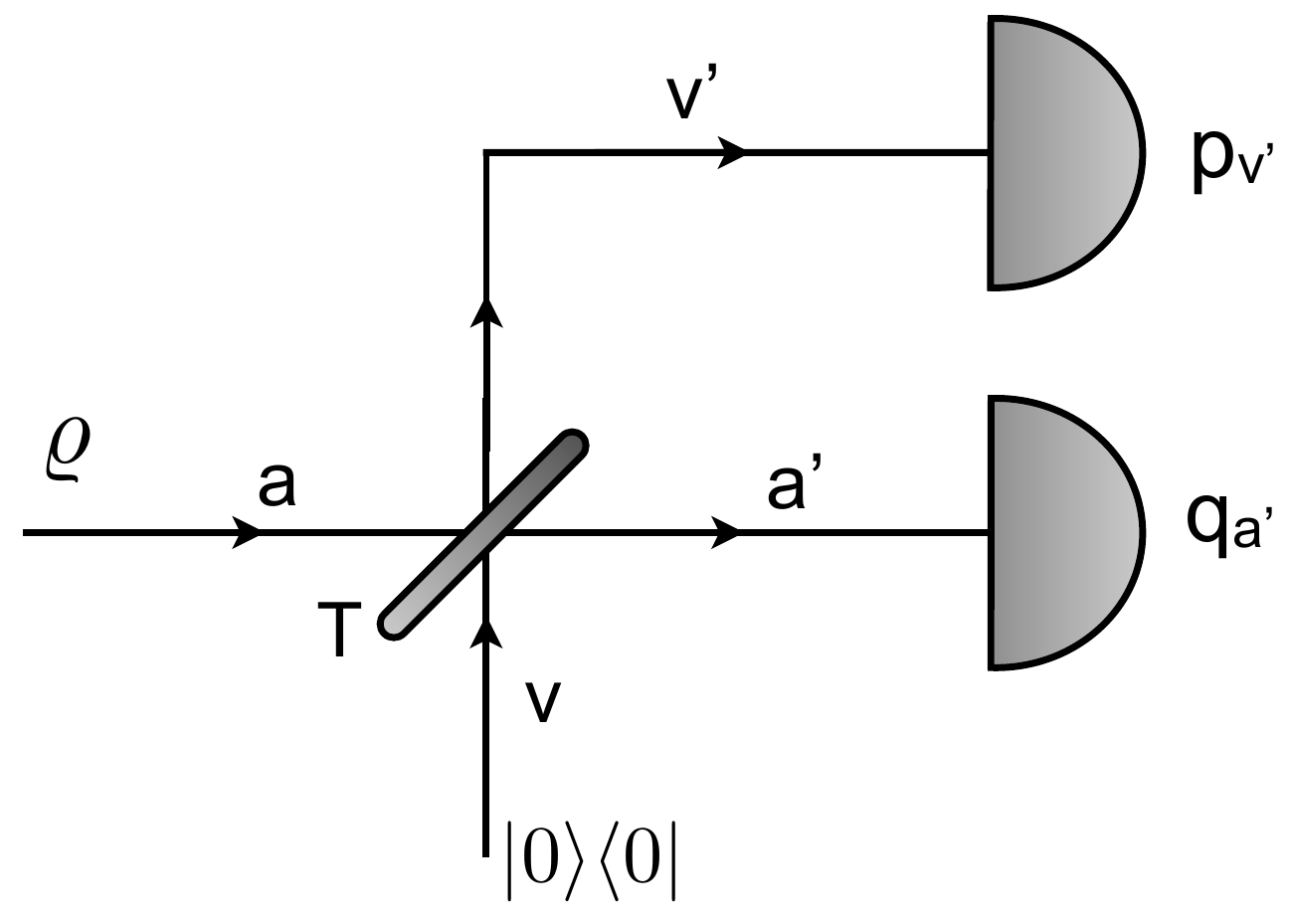}}
\caption{\label{f:doublehomodyne} Measurement scheme for general-dyne detection: the quantum state $\varrho$ (mode $a$) is coupled to the vacuum (mode $v$) at a beam splitter with transmissivity $T$. On the output modes $a'$ and $v'$, homodyne measurements of the quadratures $q_{a'}$ and $p_{v'}$ are performed.
}
\end{figure}
 \section{General-dyne measurement scheme} \label{s:scheme}
 We here present a feasible measurement scheme based on linear optics elements and homodyne detection able to measure the observable $\Theta = a  + \Upsilon a^\dag$. We start by reviewing the double-homodyne scheme depicted in Fig. \ref{f:doublehomodyne}. The quantum state to be measured 
 $\varrho$ is mixed with a vacuum state at a beam-splitter with transmissivity $T=\cos^2\phi$.
In this section the bosonic operators $a$ and $v$ will correspond respectively to the input system, prepared in the state $\varrho$, and to the ancillary system prepared in the vacuum state. Equivalently, the bosonic operators $a'$ and $v^\prime$  will describe respectively the transmitted and reflected arms. After the beam splitter interaction, a joint measurement of the two following quadrature operators is performed
 \begin{align}
 q_{a'} &= a'+a'^\dag = \cos\phi\: q_{a} -\sin\phi \: q_{v} \nonumber \\
 p_{v'} &= -i (v'-v'^\dag) = \sin\phi\: p_{a} + \cos\phi \: p_{v}. \nonumber
 \end{align}
 One can easily prove that this joint measurement is permitted as the two operators commute, {\em i.e.}
 $[q_{a'},p_{v'}] = 0$. Let us first consider the case of a balanced beam-splitter, {\em i.e.} $\phi=\pi/4$. The joint measurement of $q_{a'}$ and $p_{v'}$ corresponds to the measurement of the non-hermitian operator $Z=q_{a'} + i p_{v'} =a-v^\dag$. One can prove that the eigenstates of this operators have the form
 \begin{align}
 |z\rangle\rangle = D(z) |\mathbbm{1}\rangle\rangle
 \end{align}
 where $D(z)=\exp\{z a^\dag - z^* a\}$ is the displacement operator and 
 $|\mathbbm{1}\rangle\rangle = \sum_n |n\rangle |n\rangle$ is the un-normalized maximally entangled state, superposition of correlated Fock states $|n\rangle$. It is known that the probability of obtaining the result 
 {$z=z_1 + i z_2$} 
 %{\red [Here and in the following I've used for real and imaginary parts the same notation of Section II]} is
 \begin{align}
 p(z) &= \Tr_{av}[ \varrho \otimes |0\rangle_v {}_v\langle 0| |z\rangle\rangle\langle\langle z|] \nonumber \\
 &= \Tr_{a}[\varrho |z\rangle\langle z| ]
 \end{align}
 where $|z\rangle = {}_v\langle 0|z\rangle\rangle$ is a coherent state s.t. $a|z\rangle = z |z\rangle$. We have basically shown the well known result that (balance) double-homodyne detection correspond to the projection over coherent states $|z\rangle = D(z) |0\rangle$. 
 
In the following, inspired by this result, we will show that varying the beam-splitter transmissivity, {\em i.e.} by varying the angle $\phi$, we will implement the measure of the non-hermitian operator $\Theta$ introduced in Eq. (\ref{eq:Theta}).

By denoting with $R(\phi)=\exp\{ \phi (a^\dag v - a v^\dag) \}$ the beam-splitter operation, and with respectively $|q\rangle$ and $|p\rangle$, the eigenstates of the corresponding quadrature operators, we can write the probability of measuring the complex number $\theta$ as
 \begin{align}
 p(\theta) &= \Tr_{av}[ \varrho \otimes |0\rangle_v {}_v\langle 0| R^\dag(\phi) |q\rangle\langle q| \otimes |p\rangle\langle p| R(\phi)] \nonumber \\
 &= \Tr_{av} [\varrho \otimes |0\rangle_v {}_v\langle 0| R^\dag(\phiT) |z\rangle\rangle \langle\langle z| R(\phiT)]
 \end{align}
 where $\phiT=\phi-\pi/4$, and we have used the result shown above, {\em i.e.} that for $\phi=\pi/4$ the measurement correspond to projecting to the eigenstates $|z\rangle\rangle$. By tracing on the ancillary system we have
 \begin{align}
 p(\theta) = \langle \theta_\phi | \varrho |\theta_\phi\rangle 
 \end{align} 
 where 
 \begin{align}
 |\theta_\phi\rangle &= {}_v\langle 0 | R^\dag (\tilde{\phi}) |z\rangle\rangle \nonumber \\
&= {}_v\langle 0 | D(z \cos\phiT ) \otimes  D(z \sin \phiT ) R^\dag (\phiT) |\mathbbm{1}\rangle\rangle \:.\label{eq:eigA}
\end{align}
We want to show that by properly choosing $\phi$, the state $|\theta_\phi\rangle$ here derived is eigenstate of the operator $\Theta = a + \U a^\dag$. \\
By observing that $|\mathbbm{1}\rangle\rangle = \lim_{s\rightarrow\infty} S_2(s) |0\rangle$, 
{with $S_2(s)=\exp\{(s/2)(av-a^\dag v^\dag)\}$,}
{\em i.e.} it is a two-mode squeezed vacuum state with infinite entanglement, we can use the Gaussian formalism in order to derive a more useful parameterisation of the state $|\theta_\phi\rangle$.

The covariance matrix of the state $|\mathbbm{1}\rangle\rangle$ is
\begin{align}
\sigmaCM_0 = \lim_{s\rightarrow\infty} 
\left(
\begin{array}{c | c }
\cosh(2s) \mathbbm{1}_2 & \sinh(2s) \sigma_z \\
\hline
\sinh(2s) \sigma_z  & \cosh(2s) \mathbbm{1}_2
\end{array}
\right) \:,
\end{align}
with $\sigma_z={\rm diag}(1,-1)$. After applying the beam-splitter and displacement operations, the state  $|\psi\rangle = D(z \cos\phiT ) \otimes  D(-z \sin \phiT ) R^\dag (\phiT) |\mathbbm{1}\rangle\rangle $ is a Gaussian state characterized by a first-moment vector 
\begin{align}
{\bf X}_{in} = 
\left( 
\begin{array}{c} 
{\bf X}_{in}^{(a)} \\
{\bf X}_{in}^{(v)} 
\end{array}
\right)
\end{align}
where 
\begin{align}
{\bf X}_{in}^{(a)} =
\left( 
\begin{array}{c} 
\sqrt{2} {z_1} \cos\phiT \\
\sqrt{2} {z_2} \cos\phiT
\end{array}
\right) \:\:\:\:
{\bf X}_{in}^{(v)} =
\left( 
\begin{array}{c} 
\sqrt{2} z_R \sin\phiT \\
\sqrt{2} z_I \sin\phiT
\end{array}
\right);
\end{align}
analogously the covariance matrix can be evaluated as 
\begin{align}
\sigmaCM_1 &= S_{\phiT} \sigmaCM_0 S_{\phiT}^{\sf T} \nonumber \\
&= \left(
\begin{array}{c | c }
{\bf A}_1 & {\bf C}_1 \\
\hline
{\bf C}_1 & {\bf B}_1
\end{array}
\right) \:,
\end{align}
where 
\begin{align}
S_{\phi} = \left(
\begin{array}{c | c }
\cos\phi \mathbbm{1}_2 & -\sin\phi \mathbbm{1}_2  \\
\hline
\sin\phi  \mathbbm{1}_2 & \cos\phi \mathbbm{1}_2
\end{array}
\right) 
\end{align}
is the symplectic matrix corresponding to the beam-splitter evolution $R(\phi)$.

By partially projecting the state $|\psi\rangle$ on the vacuum state as in Eq. (\ref{eq:eigA}), one still obtains an output Gaussian states $|\theta_\phi\rangle$, whose first-moment vector and covariance matrix can be evaluated as
\begin{align}
{\bf X}_{out} &= {\bf X}_{in}^{(a)} + C_1^{\sf T} (B_1 + \mathbbm{1}_2)^{-1} {\bf X}_{in}^{(v)} \:,  
\label{eq:Xout}\\
\sigmaCM_{out} &= A_1 - C_1  (B_1 + \mathbbm{1}_2)^{-1} C_1^{\sf T}.
\end{align} 
By also taking the limit for $s$ that goes to infinity, one obtains
\begin{align}
{\bf X}_{out} &=
\left( 
\begin{array}{c} 
{z_1}/ \sqrt{T} \\
{z_2} / \sqrt{1-T}
\end{array}
\right) \:, \\
\sigmaCM_{out} &= {\rm diag}\left( \frac{1-T}{T} , \frac{T}{1-T} \right)
\end{align}
where $T=\cos^2\phi$. Being a pure Gaussian state, we can parametrize it as a displaced squeezed vacuum state $|\theta_\phi\rangle =
D(\beta) S(r) |0\rangle$, where $S(r) = \exp\{(r/2)(a^2 - a^{\dag 2}) \}$ denotes the squeezing operator, and
\begin{align}
\beta &= \frac{{z_1}}{\sqrt{2T}} + i \frac{{z_2}}{\sqrt{2(1-T)}} \:, \\
r &= \log \sqrt{\frac{1-T}{T}}.
\end{align}
If we now apply the operator $\Theta=a+\U a^\dag$ to the state, we have
\begin{align}
\Theta |\theta_\phi\rangle 
%&= (\beta + \U \beta^*) |\theta_\phi\rangle + D(\beta) S(r) (\mu a + \nu a^\dag + \Upsilon \mu a^\dag +\U \nu a) |0\rangle \nonumber \\
&= (\beta + \U \beta^*) |\theta_\phi\rangle +  (\nu + \U \mu) D(\beta)S(r)|1\rangle \label{eq:condition}
\end{align}
where we have defined $\mu=\cosh(2r)$, $\nu=\sinh(2r)$, and we have used the relation $S^\dag(r) a S(r) = \mu a + \nu a^\dag$. By observing the just derived Eq. (\ref{eq:condition}), it is clear that the state $|\theta_\phi\rangle$ is an eigenstate of the operator $\Theta$ if the condition $\nu + \U \mu = 0$
is fulfilled. One can easily show that this corresponds to choose the transmissivity of the beamsplitter as
\begin{align}
T_\U = \cos^2 \phi_\U = \frac{1+\U}{2} .
\end{align}
In particular in this case we have $\Theta |\theta_\phi\rangle = \theta |\theta_\phi\rangle$, where the eigenvalue reads 
$$
\theta = \beta + \U \beta^* = {z_1} \sqrt{1+\U} + i {z_2} \sqrt{1-\U}\:.
$$ 
We also observe, that by considering the measurement scheme presented, the following relation between the mean-values of the operators
\begin{align}
\langle \Theta \rangle &= \langle a + \U a^\dag\rangle \nonumber \\
&= \langle q_{a'} \rangle \sqrt{1+\U} + i \langle p_{v'} \rangle \sqrt{1-\U}
\end{align}
showing that, operationally, in order to obtain the desired value of observable $\Theta$, one has to measure the operators $q_{a'}$ and $p_{v'}$, and then multiply the outcomes by respectively $\sqrt{1+\U}$ and $\sqrt{1-\U}$.
\section{Conclusions and outlooks} \label{s:conclusions}
We have derived the stochastic master equation corresponding to continuous general-dyne measurements on a thermal bath interacting with a single bosonic mode. The general form of the equation 
allowed us to obtain the unravelling due to homodyne and heterodyne detection as special cases. 
Given the great interest shown in the recent years in the control of bosonic systems as in mechanical oscillators and microwave resonators, where the temperature of the bath cannot be neglected, our results will allow one to assess the usefulness of continuous measurement and feedback for cooling and quantum state engineering for these set-ups.

It would be interesting to extend our analysis to thermal unravellings 
%In \cite{Ravotto} it is indeed shown that 
{which include the possibility of monitoring ancillary modes corresponding to a (partial or complete) purification of the bath.}
\\

After acceptance of this work we became aware of Ref. \cite{ChiaWiseman}, where the authors give a completely general characterization of multi-mode general-dyne detections in terms of linear optics and homodyne measurements, including the results described in Sec.\ref{s:scheme}.
\section{Acknowledgment} 
MGG and AS acknowledges support from EPSRC through grant EP/K026267/1.

\begin{thebibliography}{99}
\bibitem{WisemanMilburn} H. M. Wiseman and G. J. Milburn, {\em Quantum Measurement and Control} (Cambridge University Press, New York, 2010).
\bibitem{opt2}S. Mancini, Phys. Rev. A {\bf 73}, 010304(R) (2006).
\bibitem{opt3}S. Mancini and H. M. Wiseman, Phys. Rev. A {\bf 75}, 012330 (2007).
\bibitem{opt4}A. Serafini and S. Mancini, Phys. Rev. Lett. {\bf 104}, 220501 (2004).
\bibitem{boundTH} M. G. Genoni, S. Mancini and A. Serafini, Phys. Rev. A {\bf 87}, 042333 (2013).
\bibitem{opt5} H. I. Nurdin and N. Yamamoto, Phys. Rev. A {\bf 86}, 022337 (2012).
\bibitem{BelavFiltering1}  V. P. Belavkin, Radiotechnika i Electronika, {\bf 25},1445 (1980).
\bibitem{Belavkin3}  V. P. Belavkin, in {\em Information, complexity, and control in
quantum physics}, edited by A. Blaquie`re, S. Dinar, and
G. Lochak (Springer, New York, 1987).
\bibitem{BelavFiltering2} V. P. Belavkin, Commun. Math. Phys., {\bf 146}, 611 (1992).
\bibitem{Barchielli} A. Barchielli, Quantum Opt. {\bf 2}, 423 (1990).
\bibitem{WisemanDiosi} H. M. Wiseman and L. Diosi, Chem. Phys. {\bf 268}, 91 (2001).
\bibitem{WisemanDoherty} H. M. Wiseman and A. C. Doherty, Phys. Rev. Lett. {\bf 94}, 070405 (2005).
\bibitem{OptoMech1} T. J. Kippenberg and K. J. Vahala, Science {\bf 321}, 1172 (2008). 
\bibitem{OptoMech2} M. Aspelmeyer, S. Groblacher, K. Hammerer and N. Kiesel, JOSA B {\bf 27}, A189 (2010).
\bibitem{cooling} J. D. Teufel {\em et al.} Nature {\bf 475}, 359 (2011); J. Chan {\em et al.}, Nature {\bf 478}, 89 (2011). 
\bibitem{FBAlessio} A. Serafini, ISRN Optics {\bf 2012}, 275016 (2012).
%\bibitem{Gardiner} C. Gardiner and P. Zoller, {\em Quantum Noise} (Springer, New York, 2010).
\bibitem{WisemanThesis} H. W. Wiseman, Ph.D. Thesis (University of Queensland, 1994).
%\bibitem{Ravotto} M. G. Genoni, S. Mancini and A. Serafini, {\em in preparation}. {\red [To be removed in my opinion]}
%\bibitem{notePOVM} It can be easily checked that $\int d\Pi(\theta)=\mathbb{I}$ by using \eqref{thx} and the completeness relation $\int dx |x\rangle\langle x|=\mathbb{I}$.
\bibitem{S01} {W. P. Schleich, {\em Quantum Optics in Phase Space}, (Wiley-VCH, Berlin, 2001)}
\bibitem{ChiaWiseman} A. Chia and H. M. Wiseman, Phys. Rev. A {\bf 84}, 012119 (2011)
\end{thebibliography}
\end{document}